\documentclass[a4paper]{jpconf}
\usepackage{amsfonts,amsmath,amssymb,amsthm,bbm}
\usepackage{graphicx,hyperref}

\newcommand{\C}{{\mathbb C}}

\newcommand{\R}{{\mathbb R}}

\newcommand{\cE}{{\mathcal E}}

\newcommand{\cH}{{\mathcal H}}

\newcommand{\SU}{\mathrm{SU}}

\newcommand{\U}{\mathrm{U}}

\newcommand{\id}{\mathbb{I}}

\def\cHNJ{\cH_N^{(J)}}
\def\cHN{\cH_N}

\newcommand{\be}{\begin{equation}}
\newcommand{\ee}{\end{equation}}
\newcommand{\beq}{\begin{eqnarray}}
\newcommand{\eeq}{\end{eqnarray}}
\newcommand{\bea}{\begin{eqnarray}}
\newcommand{\eea}{\end{eqnarray}}
\newcommand{\nn}{\nonumber}

\newcommand{\su}{{\mathfrak su}}
\renewcommand{\sl}{{\mathfrak sl}}

\renewcommand{\u}{{\mathfrak u}}

\newcommand{\la}{\langle}
\newcommand{\ra}{\rangle}

\newcommand{\f}{\frac}

\def\nn{\nonumber}
\def\pp{\partial}

\def\arr{\rightarrow}

\def\Ea{E^{(\alpha)}}
\def\Eb{E^{(\beta)}}
\def\Fa{F^{(\alpha)}}
\def\Fb{F^{(\beta)}}
\def\dag{^\dagger}

\newcommand{\matr}[2]{\left(\begin{array}{#1}#2\end{array}\right)}

\begin{document}
\title{Dynamics for a simple graph using the $\U(N)$ framework for loop quantum gravity}

\author{Enrique F. Borja$^{1,2}$, Jacobo D\'{\i}az-Polo$^3$, Laurent
Freidel$^4$, I\~{n}aki Garay$^1$, Etera R. Livine$^{4,5}$}

\address{$^1$ Institute for Theoretical Physics III, University of
Erlangen-N\"{u}rnberg, Staudtstra{\ss}e 7, D-91058 Erlangen (Germany).}
\address{$^2$Departamento de F\'{\i}sica Te\'{o}rica and IFIC, Centro Mixto Universidad de
Valencia-CSIC. Facultad de F\'{\i}sica, Universidad de Valencia,
Burjassot-46100, Valencia (Spain).}
%
\address{$^3$Department of Physics and Astronomy, Louisiana State University.
        Baton Rouge, LA, 70803-4001.}
\address{$^4$Perimeter Institute for Theoretical
Physics, 31 Caroline St N, Waterloo ON,\! Canada N2L\! 2Y5.}
%
\address{$^5$Laboratoire de Physique, ENS Lyon, CNRS-UMR 5672, 46
All\'ee d'Italie, Lyon 69007, France.}

\ead{Enrique.Fernandez@uv.es, jacobo@phys.lsu.edu,
lfreidel@perimeterinstitute.ca,
 igael@theorie3.physik.uni-erlangen.de, etera.livine@ens-lyon.fr}

\begin{abstract}
The implementation of the dynamics in loop quantum gravity (LQG) is
still an open problem. Here, we discuss a tentative dynamics for the
simplest class of graphs in LQG: Two vertices linked with an
arbitrary number of edges. We find an interesting global $\U(N)$
symmetry in this model that selects the homogeneous/isotropic
sector. Then, we propose a quantum Hamiltonian operator for this
reduced sector. Finally, we introduce the spinor representation for
LQG in order to propose a classical effective dynamics for this
model.
\end{abstract}

\section{Introduction}

Loop quantum gravity (LQG) proposes a non-perturbative mathematical
formulation of the kinematics of quantum gravity. The Hilbert space is generated by states defined over
oriented graphs whose edges are labeled by irreducible
representations of the $\SU(2)$ group and whose vertices are decorated with intertwiners
($\SU(2)$ invariant tensors). These are the
so-called spin networks. Despite the several advances that have
taken place in this field, one of the main challenges faced by the
theory is the systematic implementation of the dynamics. Our goal is to focus
on a specific model in order to propose a suitable Hamiltonian for
it.

We use the $\U(N)$ framework for $\SU(2)$-intertwiners
\cite{un1,un2,un3} to study the spin network Hilbert space of the
2-vertex graph (2 nodes joined by an arbitrary number $N$ of links)
from a new point of view. We identify a global symmetry that selects
a homogeneous and isotropic sector of this system \cite{2vertex} and
we construct the operators that leave this sector invariant. They
will be the building blocks to construct the Hamiltonian operator.

 On the other hand, the recent spinor representation for LQG
\cite{return,Freidel:2010bw,Livine:2011gp} opens a new way to study
several aspects of LQG. We apply this new formalism to the 2-vertex
graph and we propose a classical action with an interaction term which encodes the effective dynamics of this system. This interaction term is, indeed, the classical
counterpart of the quantum Hamiltonian obtained within the $\U(N)$
framework.

\section{The $\U(N)$ framework}

The $\U(N)$ framework introduced in \cite{un1,un2} is very
useful to study the Hilbert space of intertwiners with $N$ legs and to build appropriate semi-classical states \cite{un3}. The
basic tool  is the Schwinger representation
of the $\su(2)$ Lie algebra in terms of a pair of harmonic
oscillators $a$ and $b$:
$$
J_z=\f12(a\dag a-b\dag b),\quad
J_+=a\dag b,\quad J_-=a b\dag\,.
$$
%
Labeling the $N$ legs with the index $i$, we identify $\SU(2)$ invariant operators acting on pairs of (possibly equal) legs $i,j$ \cite{un1,un3}:
\be
E_{ij}=a\dag_ia_j+b\dag_ib_j, \quad (E_{ij}\dag=E_{ji}),\quad\qquad
F_{ij}=a_i b_j - a_j b_i,\quad (F_{ji}=-F_{ij}).\nn
\ee
The operators $E$ form a $\u(N)$-algebra
and they also form a closed algebra together with the operators
$F,F\dag$. Notice that the diagonal operators give the energy on
each leg, $E_{ii}=E_i$, which gives twice the spin $j_i$ of the
$\su(2)$ representation carried by that leg.
This spin $j_i$ is identified geometrically as the area associated to the leg $i$ and the total energy $E=\sum_i E_i$ gives twice the total area $J=\sum_i j_i$ associated to the intertwiner.
%
The $E_{ij}$-operators change the energy/area carried by each leg,
while still conserving the total energy, while the operators
$F_{ij}$ (resp. $F\dag_{ij}$) will decrease (resp. increase) the
total area $E$ by 2:
\be
[E,E_{ij}]=0,\qquad [E,F_{ij}]=-2F_{ij},\quad
[E,F\dag_{ij}]=+2F\dag_{ij}.\nn
\ee
The operators $E_{ij}$ allow then to navigate from state to state
within each subspace $\cHNJ$ of $N$-valent intertwiners with fixed
total area $J$; and the operators  $F\dag_{ij}$ and $F_{ij}$ allow
to go from one subspace $\cHNJ$ to the next $\cHN^{(J\pm 1)}$, thus
endowing the full space of $N$-valent intertwiners with a Fock space
structure with creation operators $F\dag_{ij}$ and annihilation
operators $F_{ij}$. Besides, it was proven \cite{un2} that each
subspace $\cHNJ$ carries an irreducible representation of $\U(N)$
generated by the $E_{ij}$ operators. Finally, it is worth pointing
out that the operators $E_{ij},F_{ij},F\dag_{ij}$ satisfy certain
quadratic constraints, which correspond to a matrix algebra \cite{2vertex}.

\section{The 2 vertex model and the quantum Hamiltonian}

\begin{figure}[h]
\begin{center}
\includegraphics[height=35mm]{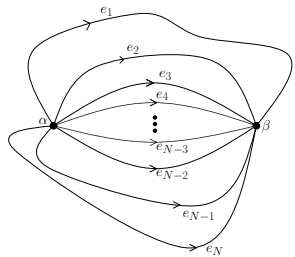}
\caption{The 2-vertex graph with vertices $\alpha$ and $\beta$ and
the $N$ edges linking them.} \label{2vertexfig}
\end{center}
\end{figure}

We consider the simplest class of non-trivial graphs for spin
network states in LQG: a graph with two vertices linked by $N$
edges, as shown in fig.\ref{2vertexfig}.

There are matching conditions \cite{un2} imposing that each edge
carries a unique $\SU(2)$ representation (same spin seen from
$\alpha$ and from $\beta$). This translates into an equal energy condition:
\be
\cE_i\,\equiv\,\Ea_i -\Eb_i \,=\,0.\nn
\ee
These constraints $\cE_k$ turn out to be part of a larger $\U(N)$
symmetry algebra. Indeed, we introduce the more
general operators:
\be
\cE_{ij}\,\equiv\, \Ea_{ij}-\Eb_{ji}
\,=\,\Ea_{ij}-(\Eb_{ij})\dag,\nn
\ee
that form a $\U(N)$ algebra and that reduce to the matching
conditions in the case $i=j$.
Now, one can show \cite{2vertex} that by imposing the global
$\U(N)$-invariance generated by the $\cE_{ij}$'s on our 2-vertex system, one obtains
a single state $|J\ra$ for each total boundary area $J$. Thus, the
$\U(N)$ invariance is restricting our system to states that are
homogeneous and isotropic (the quantum state is the same at every
point of space, i.e. at $\alpha$ and $\beta$, and all directions or
edges are equivalent).

We propose a dynamics for this system
compatible with the $\U(N)$-invariance.
Investigating the structure of the $\U(N)$-invariant operators, we propose
the most general $\U(N)$ invariant Hamiltonian (allowing only elementary
changes in the total area), up to a renormalization by a
$E$-dependent factor:
\be
H \,\equiv\, \lambda\sum_{ij}\Ea_{ij}\Eb_{ij}+ \left(\sigma
\sum_{ij}\Fa_{ij}\Fb_{ij} +\bar{\sigma}
\sum_{ij}F^{\alpha\dagger}_{ij}F^{\beta\dagger}_{ij}\right)\,,
\ee
where the coupling $\lambda$ is real while $\sigma$ can be complex a
priori, so that the operator $H$ is Hermitian.

We studied the properties of this Hamiltonian on the $\U(N)$ invariant Hilbert space. Its action on states $|J\ra$ is known and its spectral properties have been analyzed  \cite{2vertex}. It turns out that it shares several mathematical analogies with the evolution operator used in loop
quantum cosmology. At the physical level, interpreted as a cosmological model, this simple dynamical 2-vertex model also leads generically to a big bounce and avoids the big bang singularity.


\section{Spinors and effective dynamics}

Based on the Schwinger representation of $\SU(2)$ in terms of
harmonic oscillators, it is possible to give a representation of the
classical phase of LQG in terms of spinor variables
\cite{return,Livine:2011gp}. The quantization of this classical
system will lead us back to the $\U(N)$ framework for intertwiners.
Focusing on this classical system, we write an action principle with
an effective dynamics of the spinors reflecting the quantum dynamics
defined above.

Let us start by introducing the usual spinor notation. Let us define the spinor $z$ and its dual:
$$
|z\ra=\matr{c}{z^0\\z^1}\in\C^2, \qquad \la z|=\matr{cc}{\bar{z}^0
&\bar{z}^1}, \qquad |z]\equiv
\begin{pmatrix}-\bar{z}^1\\\bar{z}^0 \end{pmatrix}.
$$
In order now to describe $N$-valent intertwiners, we consider $N$
spinors $z_i$ satisfying a closure condition\footnotemark\,that, in
terms of their components, is given by:
\begin{equation}
\sum_i |z_i\ra\la z_i|\propto\id\,\Leftrightarrow\,
\sum_i z^0_i\,\bar{z}^1_i=0,\quad \sum_i \left|z^0_i\right|^2=\sum_i
\left|z^1_i\right|^2=\f12\sum_i \la z_i|z_i\ra. \label{closure}
\end{equation}
Solutions are parameterized in terms of a positive number
$\lambda\in\R_+$ and a unitary matrix $u\in\U(N)$ up to
$\U(N-2)\times \SU(2)$ right-transformations with
$z_i^0=\sqrt{\lambda}\,u_{i1}$ and $z_i^1=\sqrt{\lambda}\,u_{i2}$ .
%
\footnotetext{We associate to each spinor $z_i$ a 3-vector
$\vec{X}(z_i)=\la z_i|\vec{\sigma}|z_i\ra$ by projecting it onto the Pauli matrices.
Then the closure constraint is $\sum_i \vec{X}(z_i)=0$ and we identify $\vec{X}(z_i)$ as the normal vector to the dual surface to the leg $i$. }

The phase space is defined by the canonical Poisson bracket
$\{z^a_i,\bar{z}^b_j\}\,\equiv\,i\,\delta^{ab}\delta_{ij}\,$. The
quantization will be promoting $z_i$ and $\bar{z_i}$ as the
annihilation and creation operators of harmonic oscillators. Then
the classical matrices $M_{ij}=\la z_i |z_j \ra$ and $Q_{ij}=[z_j
|z_i\ra$ are the classical counterparts of the operators $E$ and
$F$.

The $\U(N)$-action  on spinors is the simple $N\times N$ matrix action $(Uz)_i=\sum_j U_{ij}z_j$. Defining the ``homogeneous cosmological" sector as the $\U(N)$-invariant sector, satisfying $\la z_i^\alpha |z_j^\alpha \ra=\la z_i^\beta |z_j^\beta \ra$ and invariant under $z^\alpha,z^\beta\,\arr Uz^\alpha,\bar{U}z^\beta$, imposes that all the $\alpha$-spinors are equal to the $\beta$-spinors up to a global phase, $\bar{z}^{(\alpha)}_i\,=\,e^{i\phi}\,z^{(\beta)}_i$. And we get a reduced phase space with two parameters, the total area $\lambda$ and its conjugate angle $\phi$ encoding the curvature. Our ansatz for the dynamics of this ``cosmological'' sector is:
\be
S_{inv}[\lambda,\phi] \,=\, -2 \int dt\,\left(\lambda \pp_t \phi
-\lambda^2\left(\gamma^0-\gamma^+e^{2i\phi}
-\gamma^-e^{-2i\phi}\right)\right),
\ee
which corresponds to the quantum Hamiltonian defined above. In this classical case, the equations of motion can be solved exactly \cite{return} with certain interesting analogies with (the effective dynamics of) loop quantum cosmology, showing that the dynamics of the $\U(N)$-invariant sector of the 2-vertex graph model can be interpreted as describing homogeneous and isotropic cosmology.

\section{Conclusions}

The $\U(N)$ framework and the spinor representation introduced and
studied in \cite{un1,un2,un3,2vertex,return,Freidel:2010bw,Livine:2011gp}
represents a new and refreshing way to tackle several important
issues in loop quantum gravity.

In this work we have discussed these new frameworks and we have
reviewed a proposal for the dynamics of the homogeneous and
isotropic sector of the model, both at the quantum (using the
$\U(N)$ framework) and the classical (using spinors) level. In this
process, we have described the main features of the $\U(N)$
framework, like the Fock space structure of the Hilbert space of
intertwiners with $N$ legs.
We further used this $\U(N)$ structure on the 2-vertex graph to
define a symmetry reduction to the homogeneous and isotropic sector.
We have then introduced a Hamiltonian consistent with this symmetry
reduction, which can be solved exactly and shown to be analogous
with the dynamics of loop quantum cosmology.


\section*{Acknowledgments}

This work was in part supported by the Spanish MICINN research
grants FIS2008-01980, FIS2009-11893 and ESP2007-66542-C04-01 and by
the grant NSF-PHY-0968871. IG is supported by the Department of
Education of the Basque Government under the ``Formaci\'{o}n de
Investigadores'' program.


\end{document}